**Fabio Cardone[1,2], Alessio Marrani[3] and Roberto Mignani[2-5]**

[1] Dipartimento di Fisica, Università dell'Aquila,
Via Vetoio
I - 67010 COPPITO,L'Aquila, Italy

[2] I.N.D.A.M. - G.N.F.M.

[3] Università degli Studi "Roma Tre"
Via della Vasca Navale, 84
I-00146 ROMA, Italy

[4] I.N.F.N. - Sezione di Roma III

[5] mignani@fis.uniroma3.it

# A geometrical meaning to the electron mass from breakdown of Lorentz invariance

## Abstract

We discuss the problem of the electron mass in the framework of Deformed Special Relativity (DSR), a generalization of Special Relativity based on a deformed Minkowski space (*i.e.* a four-dimensional space-time with metric coefficients depending on the energy). We show that, by such a formalism, it is possible to derive the value of the electron mass from the space-time geometry via the experimental knowledge of the parameter of local Lorentz invariance breakdown, and of the Minkowskian threshold energy $E_{0,e.m.}$ for the electromagnetic interaction. We put forward the suggestion that mass generation can be related, in DSR, to the possible dependence of mass on the metric background (*relativity of mass*).

## 1 - Introduction

The problem of the mass spectrum of the known particles (leptons and hadrons) is still an open one from the theoretical side. As a matter of fact, the Standard Model of electromagnetic, weak and strong interactions is unable to say why a given particle does bear that given (experimental) mass. As to the carriers of the four fundamental forces, symmetry considerations would require they are all massless. However, it is well known that things are not so simple: weak quanta are massive. It is therefore necessary, in the framework of the Glashow-Weinberg-Salam model of electroweak interaction, to hypothesize the Goldstone mechanism, able to give weak bosons a mass by interaction with the (till now unobserved!) Higgs boson.

On this respect, even the first, best known and familiar particle, the electron, is still a mysterious object. In spite of the successes of the Dirac equation, which allows one to The origin of its mass is far from being understood. The classical electron theory (with the works by Abraham, Lorentz and Poincaré) attempts at considering the mass of the electron as of purely electromagnetic origin, and is well known to be deficient on several respects. The basic flaw of such a picture is due to the Ernshaw theorem, stating that it is impossible to have a stationary nonneutral charge distribution held together by purely electric forces. Moreover, a purely electromagnetic model of the electron implies the occurrence of divergent quantities. Such infinities can be dealt with by means of the renormalization procedure in Quantum Electrodynamics (QED). However, even in this framework, the value of the electron mass is not the intrinsic one but only that resulting from its interaction with the vacuum.

The modern view to the problem of the electron mass is that pioneered by Wheeler and Feynman[1], according to which it is not of electromagnetic origin but entirely mechanical[2].

In this paper, we want to show that the electron mass $m_e$ can be obtained from arguments related to the breakdown of local Lorentz invariance, in the framework of a generalization of Special Relativity (*Deformed Special Relativity*, DSR), based on a "deformation" of the Minkowski space (i.e. with metric coefficients depending on the energy) . This allows one to attribute to $m_e$ a geometrical meaning, by expressing it in terms of the parameter $d$ of LLI breakdown.

The organization of the paper is as follows. In Sect. 2 we briefly introduce the concept of deformed Minkowski space, and give the explicit forms of the phenomenological energy-dependent metrics for the four fundamental interactions. The LLI breaking parameter $d_{int}$ for a given interaction is introduced in Sect. 3. In Sect. 4 we assume the existence of a stable fundamental particle interacting gravitationally, electromagnetically and weakly, and show (by imposing some physical requirements) that its mass value (expressed in terms of $d_{e.m.}$ and $E_{0,grav}$ is just the electron mass. In Sect. 5 we briefly introduce the concept of *mass relativity* in DSR. Sect. 6 concludes the paper.

## 2- Deformed Special Relativity in four dimensions (DSR)

### 2.1 - Deformed Minkowski space-time

Deformed Special Relativity is a generalization of Special Relativity (SR) based on a "deformed" Minkowski space, assumed to be endowed with a metric whose coefficients depend on the energy of the process considered[3]. Such a deformation is essentially aimed at providing a metric representation of the interaction ruling the process considered (at least in the related energy range, and *locally*, *i.e.* in a suitable space-time region)[3-6]. DSR applies in principle to *all* four interactions (electromagnetic, weak, strong and gravitational), at least as far as their non-local behavior and non-potential part are concerned.

The generalized ("deformed") Minkowski space $\tilde{M}_4$ (DMS4) is defined as a space with the same local coordinates $x$ of $M_4$ (the four-vectors of the usual Minkowski space), but with metric given by the metric tensor [1]

$$
\begin{aligned}
\boldsymbol{h}_{\boldsymbol{mn}}(E) &= diag\left(b_0^2(E), -b_1^2(E), -b_2^2(E), -b_3^2(E)\right)^{ESCoff} \equiv \\
&\equiv \boldsymbol{d}_{\boldsymbol{mn}}\left[\boldsymbol{d}_{\boldsymbol{m0}}b_0^2(E) - \boldsymbol{d}_{\boldsymbol{m1}}b_1^2(E) - \boldsymbol{d}_{\boldsymbol{m2}}b_2^2(E) - \boldsymbol{d}_{\boldsymbol{m3}}b_3^2(E)\right]
\end{aligned}
\tag{2.1}
$$

$\left(\forall E \in R_0^+\right)$, where the $\{b_\mu^2(E)\}$ are dimensionless, real, positive functions of the energy[3]. The generalized interval in $\tilde{M}_4$ is therefore given by ($x^\mu = (x^0, x^1, x^2, x^3) = (ct, x, y, z)$, with $c$ being the usual light speed in vacuum) (ESC on)

$$
\begin{aligned}
ds^2 &= b_0^2(E)c^2 dt^2 - \left(b_1^2(E)dx^2 + b_2^2(E)dy^2 + b_3^2(E)dz^2\right) = \\
&= \boldsymbol{h}_{\boldsymbol{mn}}(E)dx^{\boldsymbol{m}}dx^{\boldsymbol{n}} \equiv dx * dx.
\end{aligned}
\tag{2.2}
$$

The last step in (2.2) defines the scalar product $*$ in the deformed Minkowski space $\tilde{M}_4$ [2]. It follows immediately that it can be regarded as a particular case of a Riemann space with null curvature.

Let us stress that, in this formalism, the energy $E$ is to be understood as the energy of a physical process measured by the detectors via their electromagnetic interaction in the usual Minkowski space. Moreover, $E$ is to be considered as a dynamical variable, because

---

it specifies the dynamical behavior of the process under consideration, and, via the metric coefficients, it provides us with a dynamical map - in the energy range of interest - of the interaction ruling the given process. Let's recall that the use of momentum components as dynamical variables on the same foot of the space-time ones can be traced back to Ingraham[9] . Dirac[10], Hoyle and Narlikar[11] and Canuto et al.[12] treated mass as a dynamical variable in the context of scale-invariant theories of gravity.

Moreover, it was shown that the DSR formalism is actually a five-dimensional one, in the sense that the deformed Minkowski space can be naturally embedded in a larger Riemannian manifold, with energy as fifth dimension[13] . Curved 5-d. spaces have been considered by several Authors[14]. On this respect, the DSR formalism is a kind of generalized (*non-compactified*) Kaluza-Klein theory, and resembles, in some aspects, the so-called "Space-Time-Mass" (STM) theory (in which the fifth dimension is the rest mass), proposed by Wesson[15] and studied in detail by a number of Authors[16].

By putting $ds^2 = 0$ , we get the *maximal causal velocity* in $\widetilde{M}_4^{(3,21)}$

$$\vec{u}(E) \equiv \left( c\frac{b_0(E)}{b_1(E)}, c\frac{b_0(E)}{b_2(E)}, c\frac{b_0(E)}{b_3(E)} \right) \qquad (2.3)$$

(*i.e.* the analogous of the light speed in SR) for the interaction represented by the deformed metric considered.

In DSR the relativistic energy, for a particle of mass $m$ subjected to a given interaction and moving along $\hat{x}_i$ ,has the form[3]:

$$E = m\, u^2_i(E) \widetilde{\boldsymbol{g}}(E) == m\, c^2 \frac{b^2_0(E)}{b^2_i(E)} \widetilde{\boldsymbol{g}}(E) \qquad (2.4)$$

where $u_i$ is the i-th component of the maximal velocity (2.3) of the interaction considered, and

$$\widetilde{\boldsymbol{g}}(E) \equiv \left(1 - \widetilde{\boldsymbol{b}}^2_i\right)^{-\frac{1}{2}} = \left[1 - \left(\frac{v_i b_i(E)}{c b_0(E)}\right)^2\right]^{-\frac{1}{2}} ;$$

$$\widetilde{\boldsymbol{b}}_i \equiv \frac{v_i}{u_i} \qquad (2.5)$$

In the non-relativistic (NR) limit of DSR, i.e. at energies such that

$$v_i \div u_i(E) \qquad (2.6)$$

Eq. (2.4) yields the following NR expression of the energy corresponding to the given interaction:

$$E_{NR} = mu^2{}_i(E) = mc^2 \frac{b^2{}_0(E)}{b^2{}_i(E)}$$

<div align="right">(2.7)</div>

## 2.2 - Energy-dependent phenomenological metrics
for the four interactions

As far as phenomenology is concerned, we recall that a local breakdown of Lorentz invariance may be envisaged for all four fundamental interactions (electromagnetic, weak, strong and gravitational) whereby *one gets evidence for a departure of the space-time metric from the Minkowskian one* (at least in the energy range examined). The experimental data analyzed were those of the following four physical processes: the lifetime of the (weakly decaying) $K^0{}_S$ meson[17]; the Bose-Einstein correlation in (strong) pion production[18]; the superluminal photon tunneling[19]; the comparison of clock rates in the gravitational field of Earth[20]. A detailed derivation and discussion of the energy-dependent phenomenological metrics for all the four interactions can be found in Ref.s [3-6]. Here, we confine ourselves to recall their following basic features:

1) Both the electromagnetic and the weak metric show the same functional behavior, namely

$$\mathbf{h}_{mn}(E) = diag\left(1, -b^2(E), -b^2(E), -b^2(E)\right),$$

<div align="right">(2.8)</div>

$$b^2(E) = \begin{cases} (E/E_0)^{1/3}, & 0 < E \leq E_0 \\ 1, & E_0 < E \end{cases} =$$
$$= 1 + \boldsymbol{q}\,(E_0 - E)\left[\left(\frac{E}{E_0}\right)^{1/3} - 1\right], E > 0$$

<div align="right">(2.9)</div>

(where $?(x)$ is the Heaviside theta function) with the only difference between them being the threshold energy $E_0$, i.e. the energy value at which the metric parameters are constant, i.e. the metric becomes Minkowskian ( $?_{\mu?}(E = E_0) = g_{\mu?} = diag(1,-1,-1,-1)$); the fits to the experimental data yield

$$E_{0,e.m.} = (4.5 \pm 0.2)\ \mu eV;$$

<div align="right">(2.10)</div>

$$E_{0,weak} = (80.4 \pm 0.2)\ GeV;$$

Notice that for either interaction the metric is isochronous, spatially isotropic *and "sub-Minkowskian"*, i.e. it approaches the Minkowskian limit from below (for $E<E_0$). Both metrics are therefore Minkowskian for $E>E_{0,weak} > 80$ *GeV*, and then our formalism is fully consistent with electroweak unification, which occurs at an energy scale ~ 100 *GeV*.

Let us recall that the phenomenological electromagnetic metric (2.8)-(2.10) was derived by analyzing the propagation of evanescent waves in undersized waveguides[17]. It allows one to account for the observed superluminal group speed in terms of a nonlocal behavior of the waveguide, just described by an effective deformation of space-time in its reduced part[5]. As to the weak metric, it was obtained by fitting the data on the meanlife of the meson $K^0_S$ (experimentally known in a wide energy range (30÷350 *GeV*)[17]), thus accounting for its apparent departure from a purely Lorentzian behavior[3,21].

2) For the strong interaction, the metric was derived[4] by analyzing the phenomenon of Bose-Einstein (BE) correlation for p -mesons produced in high-energy hadronic collisions[18]. Such an approach permits to describe the BE effect as the decay of a "fireball" whose lifetime and space sizes are directly related to the metric coefficients $b^2_{\mu,strong}(E)$, and to avoid the introduction of "ad hoc" parameters in the pion correlation function[4]. The strong metric reads

$$\mathbf{h}_{strong}(E) = diag\left(b^2_{0,strong}(E), -b^2_{1,strong}(E), -b^2_{2,strong}(E), -b^2_{3,strong}(E)\right) \qquad (2.11)$$

$$
\left.
\begin{aligned}
b^2_{1,strong}(E) &= \left(\frac{\sqrt{2}}{5}\right)^2 \\
b^2_{2,strong}(E) &= \left(\frac{2}{5}\right)^2
\end{aligned}
\right\} \forall E > 0,
$$

$$b^2_{0,strong}(E) = b^2_{3,strong}(E) = \begin{cases} 1, & 0 < E \leq E_{0,strong} \\ \left(E/E_{0,strong}\right)^2, & E_{0,strong} < E \end{cases} = \qquad (2.12)$$

$$= 1 + \mathbf{q}(E - E_{0,strong})\left[\left(\frac{E}{E_{0,strong}}\right)^2 - 1\right], E > 0$$

with

$$E_{0,strong} = (\,367.5 \pm 0.4\,)\,GeV\,. \qquad (2.13)$$

Let us stress that, in this case, contrarily to the electromagnetic and the weak ones, *a deformation of the time coordinate occurs*; moreover, *the three-space is anisotropic*, with two spatial parameters constant (but different in value) and the third one variable with energy like the time one.

3) The gravitational energy-dependent metric was obtained[6] by fitting the experimental data on the relative rates of clocks in the Earth gravitational field[20]. Its explicit form is[3]:

$$\boldsymbol{h}_{grav}(E) = diag\left(b_{0,grav}^2(E), -b_{1,grav}^2(E), -b_{2,grav}^2(E), -b_{3,grav}^2(E)\right) \qquad (2.14)$$

$$b_{0,grav}^2(E) = b_{3,grav}^2(E) = \begin{cases} 1, & 0 < E \le E_{0,grav} \\ \dfrac{1}{4}\left(1 + \dfrac{E}{E_{0,grav}}\right)^2, & E_{0,grav} < E \end{cases} =$$

$$= 1 + \boldsymbol{q}(E - E_{0,grav})\left[\dfrac{1}{4}\left(1 + \dfrac{E}{E_{0,grav}}\right)^2 - 1\right], E > 0 \qquad (2.15)$$

with

$$E_{0,grav} = (20.2 \pm 0.1)\ \mu eV. \qquad (2.16)$$

Intriguingly enough, this is approximately of the same order of magnitude of the thermal energy corresponding to the 2.7°K cosmic background radiation in the Universe[4].

Notice that the strong and the gravitational metrics are *over-Minkowskian* (namely, they approach the Minkowskian limit from above ($E_0 < E$), at least for their coefficients $b_0^2(E) = b_3^2(E)$).

# 3 - LLI breaking factor in DSR

The breakdown of standard local Lorentz invariance (LLI) is expressed by the LLI breaking factor parameter $d$ [23]. We recall that two different kinds of LLI violation parameters exist: the isotropic (essentially obtained by means of experiments based on the propagation of e.m. waves, e.g. of the Michelson-Morley type), and the anisotropic ones (obtained by experiments of the Hughes-Drever type[23], which test the isotropy of the nuclear levels).

In the former case, the LLI violation parameter reads[23]

---

[3] The coefficients $b_{1,grav}^2(E)$ and $b_{2,grav}^2(E)$ are presently undetermined at phenomenological level.

[4] It is worth stressing that the energy-dependent gravitational metric (2.14)-(2.16) is to be regarded as a *local* representation of gravitation, because the experiments considered took place in a neighborhood of Earth, and therefore at a small scale with respect to the usual ranges of gravity (although a large one with respect to the human scale).

$$\boldsymbol{d} = \left(\frac{u}{c}\right)^2 - 1, \tag{3.1}$$

$$u = c + v$$

where $c$ is, as usual, the speed of light *in vacuo*, $v$ is the LLI breakdown speed (e.g. the speed of the preferred frame) and $u$ is the new speed of light (i.e. the *"maximal causal speed"* in Deformed Special Relativity[(3)]). In the anisotropic case, there are different contributions $d^{\mathrm{A}}$ to the anisotropy parameter from the different interactions. In the HD experiment, it is A=S, HF, ES, W, meaning strong, hyperfine, electrostatic and weak, respectively. These correspond to four parameters $d^{\mathrm{S}}$ (due to the strong interaction), $d^{\mathrm{ES}}$ (related to the nuclear electrostatic energy), $d^{\mathrm{HF}}$ (coming from the hyperfine interaction between the nuclear spins and the applied external magnetic field) and $d^{\mathrm{W}}$ (the weak interaction contribution).

All the above tests put upper limits on the value of $d$ [(23)].

Moreover, at the end of the past century, a new electromagnetic experiment was proposed[(24)], aimed at directly testing LLI. It is based on the possibility of detecting a non-zero Lorentz force between the magnetic field **B** generated by a stationary current $I$ circulating in a closed loop $\gamma$, and a charge $q$, in the hypothesis that both $q$ and $\gamma$ are at rest in the same inertial reference frame. Such a force is zero, according to the standard (relativistic) electrodynamics. The results obtained by such a method in two experimental runs[(25)] admit as the most natural interpretation the fact *that local Lorentz invariance is in fact broken.*

The value of the (isotropic) LLI breaking factor determined by this electromagnetic experiment is[(25)]

$$d \cong 4 \cdot 10^{-11} \tag{3.2}$$

and represents the present lowest limit to $d$.

In order to establish a connection with the electron mass, we can define the LLI breakdown parameter for a given interaction, $d_{int}$, as

$$\boldsymbol{d}_{\mathrm{int}} \equiv \frac{m_{in,\mathrm{int.}} - m_{in,grav.}}{m_{in,\mathrm{int.}}} = 1 - \frac{m_{in,grav.}}{m_{in,\mathrm{int.}}} \tag{3.3}$$

where $m_{in,\mathrm{int.}}$ is the inertial mass of the particle considered with respect to the given interaction[5] . In other words, we assume that the *local* deformation of space-time corresponding to the interaction considered, and described by the metric (2.1), gives rise to a *local violation* of the Principle of Equivalence for interactions different from the gravitational one. Such a departure, just expressed by the parameter $d_{int}$, does constitute also a measure of the amount of LLI breakdown. In the framework of DSR, $d_{int}$ embodies the *geometrical contribution to the inertial mass*, thus discriminating between two different metric structures of space-time.

---

[5] Throughout the present work, "*int*" denotes a physically detectable fundamental interaction, which can be operationally defined by means of a phenomenological energy-dependent metric of deformed-Minkowskian type.

Of course, if the interaction considered is the gravitational one, the Principle of Equivalence strictly holds, i.e.

$$m_{in.,grav.} = m_g \qquad (3.4)$$

where $m_g$ is the gravitational mass of the physical object considered, i.e. it is its *"gravitational charge"* (namely its coupling constant to the gravitational field).

Then, we can rewrite (3.3) as:

$$\boldsymbol{d}_{\text{int.}} = \frac{m_{in.,\text{int.}} - m_g}{m_{in.,\text{int.}}} = 1 - \frac{m_g}{m_{in.,\text{int.}}} \qquad (3.5)$$

and therefore, when the particle is subjected only to gravitational interaction, it is

$$d_{grav.} = 0 \qquad (3.6)$$

In the case of the gravitational metric (2.14)-(2.15), we have

$$\frac{b_{0,grav.}(E)}{b_{3,grav.}(E)} = 1, \forall E \in R_0^{\,+} \qquad (3.7)$$

Therefore, for *i*=3 , Eq. (2.4) yields, for the gravitational energy of a particle moving along the z-axis ($v_3 = v$):

$$E_{grav} = m_g c^2 \left[ 1 - \left( \frac{v}{c} \right)^2 \right]^{-1/2} = m_g c^2 \boldsymbol{g}, \qquad (3.8)$$

with non-relativistic limit (cfr. Eq. (2.7))

$$E_{grav,NR} = m_g c^2 \qquad (3.9)$$

namely, the gravitational energy takes its *standard, special-relativistic* values.

This means that the special characterization (corresponding to the choice *i*=3) of Eq.s (2.4) and (2.7) within the framework of DSR relates the gravitational interaction with SR, which is - as well known - based on the electromagnetic interaction in its Minkowskian form.

### 4 - The electron as a fundamental particle and its "geometrical" mass

Let us consider for $E$ the threshold energy of the gravitational interaction:

$$E = E_{0,grav} \qquad (4.1)$$

where $E_{0,grav}$ is the limit value under which the metric $?_{\mu?,grav}(E)$ becomes minkowskian (at least in its known components). Indeed, from Eq.s (2.14), (2.15) it follows:

$$\boldsymbol{h}_{grav}(E) = diag\left(1, -b_{1,grav}^2(E), -b_{2,grav}^2(E), -1\right)$$

$$\overset{ESCoff}{=} \boldsymbol{d}_{mn}\left[\boldsymbol{d}_{m0} - \boldsymbol{d}_{m1}b_{1,grav}^2(E) - \boldsymbol{d}_{m2}b_{2,grav}^2(E) - \boldsymbol{d}_{m3}\right], \forall E \in \left(0, E_{0,grav}\right) \qquad (4.2)$$

Notice that at the energy $E = E_{0,grav}$ the electromagnetic metric (2.8),(2.9) is Minkowskian, too (because $E_{0,grav} > E_{0,e.m.}$).

On the basis of the previous considerations, it seems reasonable to assume that the physical object (particle) $p$ with a rest energy (i.e. gravitational mass) just equal to the threshold energy $E_{0,grav}$, namely

$$E_{0,grav} = m_{g,p}c^2 \qquad (4.3)$$

must play a fundamental role for either e.m. and gravitational interaction. We can e.g. hypothesize that $p$ corresponds to the lightest mass eigenstate which experiences both force fields (i.e., from a quantum viewpoint, coupling to the respective interaction carriers, the photon and the graviton). As a consequence, $p$ must be intrinsically stable, due to the impossibility of its decay in lighter mass eigenstates, even in the case such a particle is subject to weak interaction, too (i.e. it couples to all gauge bosons of the Glashow-Weinberg-Salam group $SU(2) \times U(1)$, not only to its electromagnetic charge sector).

Since, as we have seen, for $E = E_{0,grav}$ the electromagnetic metric is minkowskian, too, it is natural to assume, for $p$:

$$m_{in,p,e.m.} = m_{in,p} \qquad (4.4)$$

namely its inertial mass is that measured with respect to the electromagnetic metric.

Then, due to the Equivalence Principle (see Eq. (3.4)), the mass of $p$ is characterized by

$$p : \begin{cases} m_{in,p,grav} = m_{g,p} \\ m_{in,p,e.m.} = m_{in,p} \end{cases} \qquad (4.5)$$

Therefore, for such a fundamental particle the LLI breaking factor (3.3) of the e.m. interaction becomes:

$$\boldsymbol{d}_{e.m.} \equiv \frac{m_{in,p.} - m_{g,p}}{m_{in,p.}} = 1 - \frac{m_{g,p.}}{m_{in,p.}}$$

$$\Leftrightarrow m_{g,p} = m_{in,p}\left(1 - \boldsymbol{d}_{e.m.}\right) \tag{4.6}$$

Replacing (4.3) in (4.6) yields:

$$E_{0,grav} = m_{in,p}\left(1 - \boldsymbol{d}_{e.m.}\right)c^2 \Leftrightarrow m_{in,p} = \frac{E_{0,grav}}{c^2}\frac{1}{1 - \boldsymbol{d}_{e.m.}} \tag{4.7}$$

Eq. (4.7) allows us to evaluate the inertial mass of $p$ from the knowledge of the electromagnetic LLI breaking parameter $d_{e.m.}$ and of the threshold energy $E_{0,grav}$ of the gravitational metric.

On account of Eq. (3.1), we can relate the lowest limit to the LLI breaking factor of electromagnetic interaction, Eq. (3.3) (determined by the coil-charge experiment), with $d_{e.m}$ as follows:

$$d = 1 - d_{e.m.} \cong 4 \cdot 10^{-11} \tag{4.8}$$

Then, inserting the value (2.16) for $E_{0,grav}$ [6] and (4.8) in (4.7), we get

$$m_{in,p} = \frac{E_{0,grav}}{c^2}\frac{1}{1 - \boldsymbol{d}_{e.m.}} \geq \frac{2 \times 10^{-5}}{4 \times 10^{-11}}\frac{eV}{c^2} = 0.5\frac{MeV}{c^2} = m_{in,e} \tag{4.9}$$

(with $m_{in,e}$ being the electron mass) where the $\geq$ is due to the fact that in general the LLI breaking factor constitutes an *upper limit* (i.e. it sets the scale *under which* a violation of LLI is expected). If experiment [25] *does indeed provide evidence* for a LLI breakdown (as it seems the case, although further confirmation is needed), Eq. (4.9) yields $m_{in,p} = m_{in,e}$. We find therefore the amazing result that *the fundamental particle $p$ is nothing but the electron $e^-$ (or its antiparticle $e^+$ [7])*. The electron is indeed the lightest massive lepton (pointlike, non-composite particle) with electric charge, and therefore subjected to gravitational, electromagnetic and weak interactions, but unable to weakly decay due to its small mass. Consequently, $e^-$ ($e^+$) shares all the properties we required for the particle $p$, whereby it plays a fundamental role for gravitational and electromagnetic interactions.

---

[6] Let us recall that the value of $E_{0,grav}$ was determined by fitting the experimental data on the slowing down of clocks in the Earth gravitational field [20]. See Ref. [6].

[7] Of course, this last statement does strictly holds only if the CPT theorem maintains its validity in the DSR framework, too. Although this problem has not yet been addressed in general on a formal basis, we can state that it holds true in the case we considered, since we assumed that the energy value is $E = E_{0,grav}$ corresponding to the Minkowskian form of both electromagnetic and gravitational metric.

## 5 – Mass relativity in DSR

The considerations carried out in the previous Sections allow us therefore to relate the electron mass to the (local) breakdown of Lorentz invariance. Such a mass would then be a *measure of the deviation of metric from the Minkowskian one.* The minimum measured mass of a particle would be related to the minimum possible metric deviation compatible with its interactions.

Such a point can be reinforced by the following argument.

The maximal causal velocity $\vec{u}$ defined by Eq. (2.3) can be interpreted, from a physical standpoint, as the speed of the quanta of the interaction locally (and phenomenologically) described in terms of a deformed Minkowski space. Since these quanta are associated to lightlike world-lines in $\widetilde{M}_4$, they must be zero-mass particles (*with respect to the interaction considered*), in analogy with photons (*with respect to the e.m. interaction*) in the usual SR.

Let us clarify the latter statement. The carriers of a given interaction propagating with the speed $\vec{u}$ typical of that interaction *are actually expected to be strictly massless only inside the space whose metric is determined by the interaction considered*. A priori, nothing forbids that such "deformed photons" *may acquire a non-vanishing mass in a deformed Minkowski space related to a different interaction.*

This might be the case of the massive bosons $W^+$, $W^-$ and $Z^0$, carriers of the weak interaction. They would therefore be massless in the space $\widetilde{M}_4(\boldsymbol{h}_{weak}(E))$ related to the weak interaction, but would acquire a mass when considered in the standard Minkowski space $M$ of SR (that, as already stressed, is strictly connected to the electromagnetic interaction ruling the operation of the measuring devices). In this framework, therefore, it is not necessary to postulate a "symmetry breaking" mechanism (like the Goldstone one in gauge theories) to let particles acquire mass. On the contrary, if one could build up measuring devices based on interactions different from the e.m. one, the photon might acquire a mass with respect to such a non-electromagnetic background.

Mass itself would therefore assume a *relative nature*, related not only to the interaction concerned, but also to the *metric background* where one measures the energy of the physical system considered. This can be seen if one takes into account the fact that in general, for relativistic particles, mass is the invariant norm of 4-momentum, and what is usually measured *is not* the value of such an invariant, but of the related energy.

### 5 - Conclusions

The formalism of DSR describes - among the others -, in geometrical terms (via the energy-dependent deformation of the Minkowski metric) the breakdown of Lorentz invariance at local level (parametrized by the LLI breaking factor $d_{int}$). We have shown that within DSR it is possible - on the basis of simple and plausible assumptions - to evaluate the inertial mass of the electron $e^-$ (and therefore of its antiparticle, the positron $e^+$) by exploiting the expression of the relativistic energy in the deformed Minkowski space $\tilde{M}_4(E)_{E \in R_0^+}$, the explicit form of the phenomenological metric describing the gravitational interaction (in particular its threshold energy), and the LLI breaking parameter for the electromagnetic interaction $d_{e.m.}$.

Therefore, *the inertial properties of one of the fundamental constituents of matter and of Universe do find a "geometrical" interpretation in the context of DSR, by admitting for local violations of standard Lorentz invariance*.

We have also put forward the idea of a *relativity of mass,* namely the possible *dependence of the mass of a particle on the metric background* where mass measurements are carried out. This could constitute a possible *mechanism of mass generation* alternative to those based on *symmetry breakdown* in Relativistic Quantum Theory.